\documentclass[12pt,preprint]{aastex}
\usepackage{natbib}
\bibpunct{(}{)}{;}{a}{}{,}
\usepackage{graphicx}
\newcommand{\HII}{H$_{\rm II}$}
\slugcomment{Accepted 14 April 2003 to ApJ Letters}

\shorttitle{NIR images of W49A: Uncovering the beast}
\shortauthors{Alves \& Homeier}

\begin{document}


\title{Uncovering the Beast: Discovery of Embedded Massive Stellar Clusters
  in W49A\altaffilmark{1}}


\author{J. Alves \& N. Homeier\altaffilmark{2}} 
\affil{European Southern Observatory, Karl-Schwarzschild Str. 2, \\ 
  85748 Garching b.  M\"unchen, Germany}


\altaffiltext{1}{Based on observations collected at the European Southern
  Observatory, La Silla, Chile}
\altaffiltext{2}{Also, University of Wisconsin, Dept. of Astronomy,
  475 N. Charter St., Madison, WI 53706, USA}

\begin{abstract}
  We present subarcsecond $J$, $H$, and $K_s$ images (FWHM $\sim$
  0.5$^{\prime\prime}$) of an unbiased $5^\prime\times5^\prime$ (16 pc
  $\times$ 16 pc) survey of the densest region of the W49 giant
  molecular cloud.  The observations reveal 4 massive stellar clusters
  (with stars as massive as $\sim$~120 M$_\odot$), the larger (Cluster
  1) about 3 pc East of the well known Welch ring of ultra-compact
  \HII\ regions.  Cluster 1 is a) extincted by at least A$_V > 20$ mag
  of foreground (unrelated and local) extinction, b) has more than 30
  magnitudes of internal inhomogeneous extinction implying that it is
  still deeply buried in its parental molecular cloud, and c) is
  powering a 6 pc diameter giant \HII\ region seen both at the NIR and
  radio continuum.  We also identify the exciting sources of several
  UC~\HII\ regions.  The census of massive stars in W49A agrees or is
  slightly overabundant when compared with the number of Lyman
  continuum photons derived from radio observations.  We argue that
  although the formation of the Welch ring could have been triggered
  by Cluster 1, the entire W49A starburst region seems to have been
  multi-seeded instead of resulting from a coherent trigger.
\end{abstract}


\section{Introduction}

Massive stars are the main suppliers of heavy elements and energy into
a galaxy's interstellar medium (ISM), play a role in regulating star
formation, and in a sense, drive galaxy evolution. Yet our knowledge
of the early phases of massive stellar evolution is rather primitive,
in part because objects for study are rare due to the combination of
small number statistics and the rapidity with which they pass through
their early stages (for a review see
\citet{churchwell99,garay99,churchwell02}).  With this in mind, the
abundance of embedded massive stars in the Galactic star-forming
region W49A marks it as a scientific gem.

W49A \citep{mezger67} is one of the brightest Galactic giant radio
H~II regions ($\sim 10^7 L_\odot$), powered by the the equivalent of
about 100 O7 stars (e.g., \citet{conti02}).  It is embedded in the
densest region of a $\sim10^6$ M$_\odot$ Giant Molecular Cloud (GMC)
extending more than $\sim100$ pc in size \citep{simon01} and is the
best Galactic analogue to the starburst phenomenon seen in other
galaxies.  The W49A star forming region lies essentially on the
Galactic plane ($l = 43.17^\circ, b = +0.00^\circ$) at a distance of
11.4$\pm$1.2 kpc \citep{gwinn92}, has $\sim$40 well studied UC~\HII\ 
regions (e.g., \citet{depree97,depree00}), each associated with at
least one stars earlier than approximately B3.  About 12 of these
radio sources are arranged in the well known Welch ``ring''
\citep{welch87}.
%A few other young Galactic clusters have a large number of massive
%stars, e.g., $\eta$ Carina \citep{walborn95,rathborne02}, NGC~3603
%\citep{brandl99}, Arches cluster \citep{serabyn98}, but no other known
%region has a high a number of massive stars in such a highly embedded
%and early evolutionary state. For this reason W49A is unique in our
%Galaxy.  Recently, \citet{conti02} conducted a H and K-band imaging
%survey covering about 1.5 arcmin$^2$ roughly centered on the Welch
%ring and identified 2 of the exciting massive stars associated with
%UC~\HII\ regions (sources F and J2).

In an attempt to uncover the embedded stellar population in W49A we
performed an unbiased $5^\prime\times5^\prime$ (16 pc $\times$ 16 pc),
deep $J$, $H$, and $K_s-$band imaging survey centered on the densest
region of the W49 GMC \citep{simon01}.  In this Letter we present the
first results of this survey, namely a previously unknown massive
stellar cluster about 3 pc East of the Welch ring of ultra-compact
\HII\ regions and still deeply embedded in the GMC, as well as 3
relatively smaller stellar clusters.

\begin{figure*}[t]
  \centering 
\caption{Subset $5^{\prime}\times5^{\prime}$ $JHKs$  color 
  composite of our survey. North is up and East is to the left. The
  red, green, and blue channels are mapped logarithmically to the
  $K_s$, $H$, and $J-$band respectively. The labels identify known
  radio continuum sources \citep{depree97}. Sources F and J2 are
  UC\HII\ regions in the Welch ring. The main cluster (Cluster 1) is
  seen NE of O3.  Several candidate exciting sources of compact \HII\ 
  regions are visible (e.g., the sources at the center of the CC, O3,
  W49A South \HII\ regions). Dark pillars of molecular material are
  seen associated with radio sources Q, and W49A South. None of W49A
  sources are optically visible. The coordinates of the image center
  are: 19:10:16.724 +09:06:11.16 (J2000).
  \label{fig:w49v3}}
\end{figure*}

\begin{figure*}[t]
  \centering 
\caption{$J$ (blue), $K_s$ (green), and 3.6 cm radio continuum (red) 
  color composite of W49A star forming region. Red only features
  represent radio \HII\ regions too embedded to be detected in our
  deep $K_s$ image (e.g., most of the Welch ring of UC \HII\ regions).
  Yellow represents features seen in both the radio continuum and
  $K_s-$band, green-only sources are $K_s$ sources too reddened to be
  detected at the $J-$band (essentially all of W49A young stellar
  population), while blue sources are foreground stars unrelated to
  the star forming region.
  \label{fig:jkradio}}
\end{figure*}

\section{Observations and Data Reduction}\label{sec:observations}

The observations were taken in June 2001, with the SofI near-infrared
camera on the ESO's 3.5m New Technology Telescope (NTT) on La Silla,
Chile, during a spell of good weather and exceptional seeing (FWHM
$\sim$ 0.5$^{\prime\prime}$).  A set of 30 dithered images of 60
seconds each were taken in the $J$, $H$, and $K_s$ filters. The images
were combined with the DIMSUM\footnote{DIMSUM is the Deep Infrared
  Mosaicing Software package developed by Peter Eisenhardt, Mark
  Dickinson, Adam Stanford, and John Ward.} package and calibrated
through observations of standard stars taken right after the program
observations and at similar airmasses.  Photometry was performed with
the DAOPHOT package in IRAF\footnote{IRAF is distributed by the
  National Optical Astronomy Observatories}.  The full details of the
data reduction will be given in a companion paper (Homeier \& Alves
2003).

\section{Results}\label{sec:results}

In Figure~\ref{fig:w49v3} we present the composite $JHK_s$ color image
for our NTT-SofI survey. The image covers an area of
$5^\prime\times5^\prime$ on the sky and the red, green, and blue
channels are mapped logarithmically to the $K_s$, $H$, and $J-$band
respectively.  Because most field stars are essentially colorless in
the near-infrared \citep[e.g.][]{alves98} one expects the color of a
star in this image to be, to first order, a qualitatively measure of
the amount of extinction towards this star. Hence, all blue stars in
this Figure are foreground sources to the star forming region.

In Figure~\ref{fig:jkradio} we present a 3.6 cm radio continuum (red),
$K_s$ (green), and $J$ (blue) color composite of the central regions
of the survey.  The radio continuum data is taken from
\citet{depree97} and has a spatial resolution of 0.8$^{\prime\prime}$,
close to the spatial resolution of the NTT images. The red-only
features in this image represent regions of ionized hydrogen so deeply
embedded in the W49A molecular cloud that they cannot be detected in
our $K_s-$band image, e.g., the Welch ring of UC \HII\ regions with
the exception of sources F and J2 (that appear in yellow in the
image). Several \HII\ regions and UC \HII\ regions detected at radio
continuum wavelengths are clearly detected on the $K_s-$band,
suggesting that the $K_s$ extended emission is dominated by hydrogen
lines. The most prominent are identified in Figure~\ref{fig:w49v3}
following \citet{depree97} nomenclature. Several point sources lie
prominently in the center of some of these regions (e.g., CC, O3, W49A
south), are extincted by A$_V > 24$ mags of visual extinction (see
below), and are excellent candidates to the exciting sources powering
these \HII\ regions \citep{homeier03}.

The main feature in Figures~\ref{fig:w49v3} and \ref{fig:jkradio} is
the central 6~pc diameter \HII\ region E of the ring of radio sources,
with a stellar cluster at its projected center. From here on we will
refer to this cluster as Cluster 1. Note that only the North part of
this 6~pc \HII\ region is visible in the $JHK_s$ color composite,
suggesting that there is a larger optical depth towards the South
of Cluster 1, perhaps due to chance alignment of the embedded compact
\HII\ regions (e.g., JJ, O3) in front of it.

Recently, \citet{conti02} conducted a $H$ and $K-$band imaging survey
covering about 1.5 arcmin$^2$ roughly centered on the Welch ring, and
were able to identify 2 of the exciting massive stars associated with
UC~\HII\ regions (sources F and J2). They also identified several
candidate O stars at the edges of their image, and speculated that
star formation may have begun on the periphery of the UC~\HII\ 
concentration. Here we can clearly see that these sources are
associated with Cluster 1, at the center of a giant \HII\ region in
Figure~\ref{fig:jkradio}, and with Clusters 3 and 4 to the SW.

We present in Figure~\ref{fig:panel} the spatial distribution of the
detected sources as a function of ($H-K_s$) color.  In
Figure~\ref{fig:cmd} we present the $H-K_s$ vs. $K_s$ Color--Magnitude
diagram for our survey. The solid line represents a 1 Myr old
population taken from the Geneva tracks \citep{lejeune01} and the
slanted dotted lines represent a reddening in this diagram of A$_V =
48$ mag. The black circles represent sources likely associated with
the new clusters (see Figure~\ref{fig:panel}).

\section{Discussion}\label{sec:discussion}

\subsection{Spatial distribution of reddened sources}
\label{sec:spat-distr-redd}

Since the W49A star forming region is at a distance of 11.4 kpc,
virtually on the Galactic plane, one expects a large amount of
unrelated line-of-sight dust extinction to W49A, as well as dust
associated with the star forming region.
%From a $^{13}$CO survey of
%the region, \citet{simon01} concluded that there is a foreground,
%non-star forming and less dense molecular cloud (GRSMC 43.30-0.33)
%along the same line-of-sight to the W49A giant molecular cloud, and
%derived a kinematical distance to this interloper cloud of about 3
%kpc.
We will take advantage of the large amounts of dust extinction to
isolate a reliable stellar population associated with W49A giant
molecular cloud. We present in Figure~\ref{fig:panel} the spatial
distribution of the detected sources as a function of ($H-K_s$) color.
Starting with the bluer sources, ($H-K_s < 1$ mag) we find a
non-uniform distribution where about $\frac{2}{3}$ of the sources are
found on the southern half of the field. Sources in this first bin are
mainly foreground sources to the W49A star forming region, extincted
by less than about 14 magnitudes of visual extinction, and the
non-uniform distribution is likely cause by an intervening cloud at a
distance of about 3 kpc (cloud GRSMC 43.30-0.33, \citep{simon01}). The
second panel ($1 < H-K_s < 1.5$ mag; $14 <$ A$_V$ $< 24$ mag) further
suggests this interpretation.  We see the opposite spatial
distribution with an increase in extinction and the region that in the
first bin seemed under-populated is now over-populated. The majority
of these stars are likely to be highly reddeneded stars in the
background of GRSMC 43.30-0.33 but further work would have to be done
to confirm this. In the third panel we clearly detect 4 clusterings of
reddened sources.  These make up the stellar population of W49A and
some are still visible in the fourth panel where we find sources
extincted by over 32 magnitudes of visual extinction, more than half
associated with the newly found clusters. The positions of these 4
clusters are given in Table 1.
      
\subsection{Star formation in the W49A Starburst Region}

Based on $H-K_s$ color and $K_s$ magnitude, we preliminarily identify
around 100 O-stars candidates associated with the W49A region, among
which $\sim$30 within the 6~pc diameter region defined by the ionized
bubble central to our images(Cluster 1), and $\sim10$ in each of the
smaller clusters to the South (clusters 2, 3, and 4).  We should bear
in mind that due to the very high values of foreground and local
(inhomogeneous) extinction we are not complete even for the most
luminous stars in W49A.  Nevertheless, this number compares
surprisingly well with the luminosity of the entire region (about
$10^{51}$ Lyman continuum photons emitted per second; Smith et al.
1978, Gwinn et al. 1992) or the equivalent of about 100 O7 V stars
\citep{vacca94}. We can now say that the census of massive stars in
W49A agrees with the number of Lyman continuum photons derived from
radio observations.  In fact, photon leakage or absorption by dust
could be operating in the W49A region, as our count of candidate
O-stars (incomplete due to severe inhomogeneous extinction), added to
the number of known UC~\HII\ regions, gives $\sim 140$ stars with
masses greater than $15-20$~M$_{\odot}$, suggestive perhaps of a
slight overabundance of ionizing stars.

It is remarkable that the Welch ring of UC\HII\ regions is seen in
projection against the edge of the giant \HII\ region powered by
Cluster 1 (Figure~\ref{fig:jkradio}).  It necessarily invokes the
classical triggering scenario of \citet{elmegreen77} as one can easily
imagine that, chronologically, the densest part of the W49A GMC
collapsed to form Cluster 1, and the combined action of stellar winds
and UV radiation compressed the abundant nearby gas to the West,
triggering its collapse into the Welch ring.  Recent theoretical
calculations \citep{mckee02} suggest timescales for massive star
formation of the order of $\sim10^5$yr which for a typical sound speed
of $\sim$10 kms$^{-1}$ agrees well with the crossing time in the Welch
ring.  However, we argue that the formation of the three smaller
clusters to the South, undoubtedly associated with the burst of star
formation in W49A, is unlikely to have been triggered by Cluster 1.
The minimum distance (because of projection) between these less
massive clusters and Cluster 1 is $\sim$6 pc, which is larger than the
bubble of ionized gas surrounding it, the giant central \HII\ region
in Figure~\ref{fig:jkradio}. Also, given the short lifetimes of
compact \HII\ regions \citep{churchwell99} and the fact that they can
be found almost over the entire surveyed region (e.g., the projected
distance between source CC and W49A South is $\sim$11 pc) suggests a
multi-seeded, largely coeval, star formation episode in the W49A.

Finally, this work suggests that star formation in W49A, the most
massive and youngest known star forming region in the Galaxy, began
earlier and extends over a larger area than previously thought.
Moreover, star formation in W49A is still ongoing (the GMC is not
exhausted yet) as 6 hot cores (the precursors of UC \HII\ regions)
were recently found in the vicinity of the Welch ring
\citep{wilner01}.  Our results show that a considerable part of the
stellar population of this Galactic starburst is accessible in the
2$\mu$m window.  Further characterization of the embedded population
(via $H$ and $K-$band spectra and adaptive optics techniques) is
called for and will surely provide much needed information on the
starburst phenomenon seen across the Universe.

%The W49A star forming region provides us with some of the youngest massive
%(proto) stars in the Milky Way.  
%No other known star forming region in
%the Galaxy is as massive and embedded (young) as W49A.  Even if star
%formation had ceased in W49A this region would already be one of the
%most massive in the Galaxy, rivalling to massive clusters as the
%Arches cluster \citep{serabyn98}. blabla...

\begin{deluxetable}{crrc}
\tabletypesize{\scriptsize}
\tablecaption{W49A stellar clusters. \label{tab:clusters}}
\tablewidth{0pt}
\tablehead{
\colhead{Cluster} & \colhead{RA (J2000)}   & \colhead{Dec (J2000)}   &
\colhead{Assoc. radio source} 
}
\startdata
1 &19:10:17.5 &$+$9:06:21 &extended    \\
2 &19:10:21.9 &$+$9:05:04 &W49A South  \\
3 &19:10:11.9 &$+$9:05:28 &S           \\
4 &19:10:10.8 &$+$9:05:14 &Q           \\

\enddata

% %% Text for table notes should follow after the \enddata but before
% %% the \end{deluxetable}. Make sure there is at least one \tablenotemark
% %% in the table for each \tablenotetext.

% \tablenotetext{a}{Sample footnote for table~\ref{tbl-1} that was generated
% with the deluxetable environment}
% \tablenotetext{b}{Another sample footnote for table~\ref{tbl-1}}

% \tablecomments{Occasionally, authors wish to append a short
% paragraph of explanatory notes that pertain to the entire table, but
% which are different than the caption.  Such notes should be placed in
% a {\tt tablecomments} command like this.}

\end{deluxetable}

\begin{figure*}
  \centering
  \caption{Spatial distribution of detected sources as a function of
    $(H-K_s)$ color. The clusters (labeled 1, 2, 3, and 4) become
    apparent in the third panel (where A$_V \sim 20$ mag). The
    non-uniform distribution of sources in panel 1 and 2 could be due
    to intervening cloud GRSMC 43.30-0.33 located at a distance of
    $\sim$3 kpc. The field showed is the same as in Figure 1.
    \label{fig:panel}}
\end{figure*}

\begin{figure*}[t]
  \centering
\caption{$(H-K_s)$ vs. $K_s$ Color-Magnitude diagram for our
  survey.  The solid line represents a 1 Myr old population taken from
  the Geneva tracks \citep{lejeune01} and the slanted lines represent
  a reddening of A$_V = 48$ mag. The black circles identify stars
  likely associated with the W49A clusters. The 90\% completeness
  limit for a star with errors less than 15\% is marked as a bold grey
  line.  
  \label{fig:cmd}}
\end{figure*}

\section{Summary}\label{sec:summary}

The main results can be summarized as follows:

1) A deep $5^\prime\times5^\prime$ NIR survey of the W49 star forming
region reveals the presence of a massive stellar cluster about
1$^{\prime}$ (3 pc in projection) to the East of the well known Welch
ring of ultra-compact \HII regions.  About 2$^{\prime}$ (6 pc) to the
Southeast of the main cluster we find a smaller cluster associated
with W49S and about 2$^{\prime}$ (6 pc) to the Southwest we find two
smaller clusters associated with radio continuum sources S and Q.
  
  2) We find more than 100 O-stars candidates in the entire survey,
  mostly associated with the stellar clusters. We are able to identify
  the likely exciting sources of well known compact and ultra-compact
  \HII\ regions, and a companion paper will follow with these results.
  The census of massive stars in W49A agrees or is slightly
  overabundant when compared to the number of Lyman continuum photons
  derived from radio observations.
  
  3) We argue that although the formation of the Welch ring of
  UC~\HII\ regions could have been triggered by the interaction of the
  main cluster with the densest regions of the giant molecular cloud,
  the entire W49A starburst region seems to have been multi-seeded
  instead of resulting from a coherent trigger.

\bibliographystyle{apj}
\bibliography{my}

\begin{thebibliography}{17}
\expandafter\ifx\csname natexlab\endcsname\relax\def\natexlab#1{#1}\fi

\bibitem[{{Alves} {et~al.}(1998){Alves}, {Lada}, {Lada}, {Kenyon}, \&
  {Phelps}}]{alves98}
{Alves}, J., {Lada}, C.~J., {Lada}, E.~A., {Kenyon}, S.~J., \& {Phelps}, R.
  1998, \apj, 506, 292

\bibitem[{{Churchwell}(1999)}]{churchwell99}
{Churchwell}, E. 1999, in NATO ASIC Proc. 540: The Origin of Stars and
  Planetary Systems, eds. C. Lada \& N. Kylafis, Kluwer, 515

\bibitem[{{Churchwell}(2002)}]{churchwell02}
{Churchwell}, E. 2002, \araa, 40, 27

\bibitem[{{Conti} \& {Blum}(2002)}]{conti02}
{Conti}, P.~S. \& {Blum}, R.~D. 2002, \apj, 564, 827

\bibitem[{{De Pree} {et~al.}(1997){De Pree}, {Mehringer}, \& {Goss}}]{depree97}
{De Pree}, C.~G., {Mehringer}, D.~M., \& {Goss}, W.~M. 1997, \apj, 482, 307

\bibitem[{{De Pree} {et~al.}(2000){De Pree}, {Wilner}, {Goss}, {Welch}, \&
  {McGrath}}]{depree00}
{De Pree}, C.~G., {Wilner}, D.~J., {Goss}, W.~M., {Welch}, W.~J., \& {McGrath},
  E. 2000, \apj, 540, 308

\bibitem[{{Elmegreen} \& {Lada}(1977)}]{elmegreen77}
{Elmegreen}, B.~G. \& {Lada}, C.~J. 1977, \apj, 214, 725

\bibitem[{{Garay} \& {Lizano}(1999)}]{garay99}
{Garay}, G. \& {Lizano}, S. 1999, \pasp, 111, 1049

\bibitem[{{Gwinn} {et~al.}(1992){Gwinn}, {Moran}, \& {Reid}}]{gwinn92}
{Gwinn}, C.~R., {Moran}, J.~M., \& {Reid}, M.~J. 1992, \apj, 393, 149

\bibitem[{{Homeier} \& {Alves}(2003)}]{homeier03}
{Homeier}, N. \& {Alves}, J. 2003, \aap\ in prep.

\bibitem[{{Lejeune} \& {Schaerer}(2001)}]{lejeune01}
{Lejeune}, T. \& {Schaerer}, D. 2001, \aap, 366, 538

\bibitem[{{McKee} \& {Tan}(2002)}]{mckee02}
{McKee}, C.~F. \& {Tan}, J.~C. 2002, \nat, 416, 59

\bibitem[{{Mezger} {et~al.}(1967){Mezger}, {Schraml}, \& {Terzian}}]{mezger67}
{Mezger}, P.~G., {Schraml}, J., \& {Terzian}, Y. 1967, \apj, 150, 807

\bibitem[{{Simon} {et~al.}(2001){Simon}, {Jackson}, {Clemens}, {Bania}, \&
  {Heyer}}]{simon01}
{Simon}, R., {Jackson}, J.~M., {Clemens}, D.~P., {Bania}, T.~M., \& {Heyer},
  M.~H. 2001, \apj, 551, 747

\bibitem[{{Vacca}(1994)}]{vacca94}
{Vacca}, W.~D. 1994, \apj, 421, 140

\bibitem[{{Welch} {et~al.}(1987){Welch}, {Dreher}, {Jackson}, {Terebey}, \&
  {Vogel}}]{welch87}
{Welch}, W.~J., {Dreher}, J.~W., {Jackson}, J.~M., {Terebey}, S., \& {Vogel},
  S.~N. 1987, Science, 238, 1550

\bibitem[{{Wilner} {et~al.}(2001){Wilner}, {De Pree}, {Welch}, \&
  {Goss}}]{wilner01}
{Wilner}, D.~J., {De Pree}, C.~G., {Welch}, W.~J., \& {Goss}, W.~M. 2001,
  \apjl, 550, L81

\end{thebibliography}

\acknowledgments

We are pleased to acknowledge Miguel Moreira for discussions and
assistance with the observations, Robert Simon for providing molecular
line data from the BU-FCRAO Galactic Ring Survey on W49A's giant
molecular cloud, where the clusters are embedded, and Chris De Pree
for providing radio continuum data of the \HII\ regions associated
with W49A.

% \begin{figure}
% \plottwo{f2a.eps}{f2b.eps}
% \caption{This is an example of a multipart figure with a long figure caption 
% that must be set as a paragraph.  The processor has to buffer the text of the
% caption, so it is good not to be too wordy, but that would make for
% poor communication as well.\label{fig2}}
% \end{figure}

%% If you use the table environment, please indicate horizontal rules using
%% Do not put multiple tabular environments within a single table.
%% The optional \label should appear inside the \caption command.

\end{document}